\documentclass[10pt,journal,compsoc]{IEEEtran}

\ifCLASSOPTIONcompsoc
  \usepackage[nocompress]{cite}
\else
  \usepackage{cite}
\fi
\usepackage{cite}
\usepackage{graphicx}
\usepackage{hhline}
\usepackage{acro}
\usepackage[switch]{lineno}
\usepackage{amssymb}
\usepackage[long,c2]{optidef}
\usepackage{booktabs}
\usepackage{bm}
\usepackage{multirow}
\usepackage{ulem}
\usepackage{tabu}
\usepackage{makecell}
\usepackage{xcolor, soul}
\sethlcolor{yellow}
\usepackage{tabularx}
\usepackage{threeparttable}

\usepackage{mathtools}

\definecolor{LightCyan}{rgb}{0.88,1,1}
\ifCLASSINFOpdf
\else
\fi
\usepackage{amsmath}
\usepackage{cuted}
\usepackage{hyperref}
\usepackage{array}
\begin{document}

\title{Multimodal-Boost: Multimodal Medical Image Super-Resolution using Multi-Attention Network with Wavelet Transform}
%
%
%
%

\author{Fayaz Ali Dharejo,~\IEEEmembership{Member,~IEEE,}
        Muhammad~Zawish, Farah Deeba, Yuanchun Zhou, Kapal Dev~\IEEEmembership{Member,~IEEE,} Sunder Ali Khowaja, and Nawab Muhammad Faseeh Qureshi,~\IEEEmembership{Senior Member,~IEEE}
\IEEEcompsocitemizethanks{\IEEEcompsocthanksitem  F.A Dharejo, F. Deeba, Y. Zhou. Authors are with the Computer Net-work Information Center, Chinese Academy of Sciences, University of Chinese Academy of Sciences, Beijing, 100190, China. E-mail: fayazdharejo@cnic.cn, deeba@cnic.cn.\protect\\
\IEEEcompsocthanksitem M. Zawish is with Walton Institute for Information and Communication Systems Science, Waterford Institute of Technology, Ireland,        E-mail: muhammad.zawish@waltoninstitute.ie
\IEEEcompsocthanksitem K. Dev is with Department of institute of intelligent systems, University of Johannesburg, South Africa.        E-mail: kapal.dev@ieee.org
\IEEEcompsocthanksitem S.A. Khowaja is with Faculty of Engineering and Technology, University of Sindh, Jamshoro, Pakistan.        E-mail: sandar.ali@usindh.edu.pk
\IEEEcompsocthanksitem N.M.F. Qureshi is associated with the Department
of Computer Education, Sungkyunkwan University, Seoul, Korea.        E-mail: faseeh@skku.edu
(Correspond: Y. Zhou E-mail: zyc @cnic.cn and  N.M.F Qureshi E-mail:faseeh@skku.edu)
}

\thanks{Manuscript received.}}

%
%

\markboth{IEEE TRANSACTIONS ON JOURNAL NAME,  MANUSCRIPT ID, September~2021}%
{Shell \MakeLowercase{\textit{et al.}}: Bare Advanced Demo of IEEEtran.cls for IEEE Computer Society Journals}
%



\IEEEtitleabstractindextext{%

\begin{abstract}
Multimodal medical images are widely used by clinicians and physicians to analyze and retrieve complementary information from high-resolution images in a non-invasive manner. Loss of corresponding image resolution adversely affects the overall performance of medical image interpretation. Deep learning based single image super resolution (SISR) algorithms has revolutionized the overall diagnosis framework by continually improving the architectural components and training strategies associated with convolutional neural networks (CNN) on low-resolution images. However, existing work lacks in two ways: i) the SR output produced exhibits poor texture details, and often produce blurred edges, ii) most of the models have been developed for a single modality, hence, require modification to adapt to a new one. This work addresses (i) by proposing generative adversarial network (GAN) with deep multi-attention modules to learn high-frequency information from low-frequency data. Existing approaches based on the GAN have yielded good SR results; however, the texture details of their SR output have been experimentally confirmed to be deficient for medical images particularly. The integration of wavelet transform (WT) and GANs in our proposed SR model addresses the aforementioned limitation concerning textons. While the WT divides the LR image into multiple frequency bands, the transferred GAN uses multi-attention and upsample blocks to predict high-frequency components. Additionally, we present a learning method for training domain-specific classifiers as perceptual loss functions. Using a combination of multi-attention GAN loss and a perceptual loss function results in an efficient and reliable performance. Applying the same model for medical images from diverse modalities is challenging, our work addresses (ii) by training and performing on several modalities via transfer learning. Using two medical datasets, we validate our proposed SR network against existing state-of-the-art approaches and achieve promising results in terms of SSIM and PSNR.
\end{abstract}

\begin{IEEEkeywords}
Super-resolution, attention modules, Multimodality data, Transfer learning, Wavelet transform, Generative adversarial network 
\end{IEEEkeywords}}

\maketitle

\IEEEdisplaynontitleabstractindextext

%
\IEEEpeerreviewmaketitle

\ifCLASSOPTIONcompsoc
\IEEEraisesectionheading{\section{Introduction}\label{sec:introduction}}
\else
\section{Introduction}
\label{sec:introduction}
\fi

%
%
%
%
\IEEEPARstart{M}{edical} images of high quality are critical in early, fast and accurate diagnosis in the current clinical processes. Typically, spatial resolution of medical images is degraded by factors such as imaging modalities and acquisition time. To recover the distorted resolution, super-resolution (SR) methods are widely adopted on digital images as post-processors eluding the additional scanning costs \cite{one}. Image super-resolution is one of the prominent low-level vision problems in the computer vision domain. The aim of image SR is to reconstruct a high resolution (HR) image from a corresponding low resolution (LR) image. SR techniques are applied on both single and multiple images simultaneously subject to the input and output criteria. In this work, we study the Single-Image-Super-Resolution (SISR) for the application of medical images. In contrast to classical SISR applications, medical imaging SISR is captious as it is often followed by precise tasks of segmentation, classification or diagnosis \cite{two} \cite{deeba2020sparse}. Thus, it becomes imperative to produce methods which could not only preserve sensitive information but also magnify the structures of interest efficiently.

\begin{figure}[h]
\includegraphics[width=0.5\textwidth]{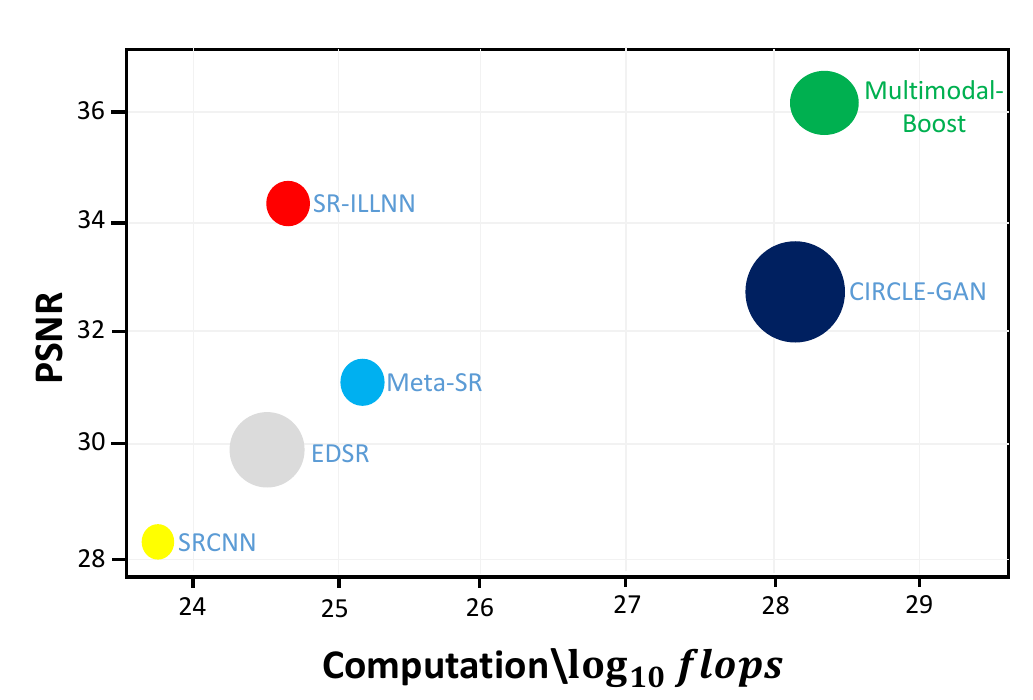}
\caption{A quantitative comparison of the super resolution capabilities of several state-of-art methods. In terms of PSNR, our proposed technique, Multimodal-Boost, exceeds the competition. The horizontal axis, which is measured in logarithmic FLOPs for easier observation and PSNR is represented on the vertical axis. The amount of parameters is represented by the circular area. A green circle in the top centered is regarded to be a superior SR network.}
\centering\label{Fig1}
\end{figure}

Traditional SR based techniques such as interpolation \cite{three} or reconstruction\cite{four} based could not suffice for medical image SISR tasks. Interpolation methods such as linear, bicubic \cite{three} and lancoz \cite{duchon1979lanczos} often fail to reconstruct the high-frequency information which eventually results in blurred and smooth edges. In contrast, reconstruction based algorithms efficiently utilize prior local\cite{tai2010super}, global\cite{yang2010exploiting}, and sparse\cite{dong2016hyperspectral} knowledge to efficiently reconstruct the HR image. However, these methods can not efficiently simulate the non-linear transformation from LR space to HR space in dynamic scenes, hence the output SR image is distorted.  Deep learning techniques have recently shown notable performance on various vision tasks such as image dehazing, object detection, activity recognition etc \cite{deebanovel}. Convolutional Neural Networks (CNNs) and particularly Generative Adversarial Aetworks (GANs) have shown remarkable performance on various SR applications such as remote sensing and medical imaging. Deep Convolutional Neural Network for Super-Resolution (SRCNN)\cite{dong2014learning} laid the foundation of deep learning (DL) based SR methods, followed by a number of techniques involving powerful capabilities of CNNs. Due to phenomenal performance of residual blocks and dense blocks, several works such as, Enhanced Deep Residual Networks for Single Image Super-Resolution (EDSR)\cite{lim2017enhanced} Very Deep Convolutional Networks (VDSR)\cite{kim2016accurate},Fast Super-Resolution with Cascading Residual Network (RFASR)\cite{ahn2018fast} and Deep Back-Projection Networks for Super-Resolution (DBPN)\cite{haris2018deep}  were proposed for the SR problem. These end to end models are deeper and complex thus requiring a large number of LR and corresponding HR pairs to get trained\cite{wang2020deep}. Moreover, these deep networks could not provide photo realistic results despite the high performance of Structural Similarity Index Measure (SSIM) and Peak Signal-to-Noise Ratio (PSNR)\cite{two}. Thus, GAN based networks accompanied by attention mechanisms\cite{xu2021multi} were introduced to produce perceptually realistic outputs closer to the ground truth HR image. However, aforementioned deep learning based SR techniques have shown sub-optimal performance on medical images particularly for different types of modalities. Several medical imaging analysis tasks such as tumor segmentation\cite{zawish2018brain} and lesion detection\cite{zhu2019lesion} are essentially hindered by the substandard HR output with blurred edges. Using a Multi-Temporal Ultra-Dense Memory Network, Peng Ye et al.\cite{yi2019multi} proposed a new method based on video super-resolution. Previous video SR methods mostly use a single memory module and a  single channel structure, which does not fully capture the correlations between frames specific to video. To achieve video  super-resolution, this paper presents a multi-temporal ultra-dense. Yiqun Mei et al.\cite{mei2020image}  developed an efficient SR method based on  Cross-Scale Non-Local Attention Module (CS-NL) incorporating recurrent neural networks to handle an inherent property of images: cross-scale correlation of features. Afterward, Kui Jiang et al.\cite{jiang2020hierarchical}  developed a hierarchical dense recursive network with super-resolution. Feature representations were made richer thanks to dense hierarchical connections. Additionally, it facilitates the lightweight SR model by providing Reasonable design and parameter sharing.

In this work, we aim to overcome aforementioned problems in multimodal medical imaging SISR by proposing a combination of multi-attention GAN with wavelet subbands. Wavelet transform (WT)\cite{abbate1995wavelet} has a unique ability of extracting useful multi-scale features with the help of sparse subbands. Since WT accompanied with GAN has been widely adopted for several applications including remote sensing for spatio-temporal SR\cite{dharejo2021twist} and dehazing\cite{fu2021dw}, it would be interesting to utilize its potential for the medical imaging SR. In the first step, we derive the WT based four subbands of the given LR image using a 2D discrete wavelet transform (DWT) from Haar family of wavelet which eventually replaces LR input with low dimensional input space preserving the key information. These subbands are then fed into the proposed GAN involving multiple attention and upsample blocks which produces an improved HR wavelet component for each corresponding subband. Lastly, the HR image is reconstructed using 2 dimensional inverse discrete wavelet transform (2D-IDWT). Inspired from this\cite{johnson2016perceptual}, we use a perceptual loss network along with GAN to further boost the performance of our proposed method in term of qualitative and qualitative. Moreover, a challenging and time-consuming task in medical imaging SR is to modify the models designed for one modality to fit a new modality. We overcome this by introducing the transfer learning technique which efficiently adapts to unseen data with the help of pre-learned knowledge. In result, it becomes feasible to adapt on datasets of different modalities along with the reliable results. In summary, the following are the contributions of our methodology:

\begin{itemize}
	\item {We proposed a novel wavelet-based multimodal and multi-attention GAN framework for medical imaging SR called Multimodal-Boost. The Multimodal-Boost can learn end-to-end residual mapping between low and high resolution wavelet subbands. The key advantage of utilizing wavelet transform is that it helps restoring rich content from images by accurately extracting missing details. WT give high-frequency information in a variety of directions that are most likely to result from (horizontal, vertical, and diagonal edges).\\}

\item {We trained the perceptual network on VGG-16 to improve the SR results, as shown in the Figure \ref{Fig7}. Using perceptual loss, we boost network and improve super-resolution performance. We use a transfer-learning technique to deal with inadequate training for super-resolution medical applications. Our algorithm is trained on DIV2K \cite{timofte2017ntire} dataset prior to getting evaluated on multimodal medical datasets.\\}

\item	{Unlike many previous \hl Deep Neural networks (DNN) medical image SR approaches \cite{zhang2017deep} \cite{chen2021super} \cite{deeba2020wavelet}\cite{romano2016raisr}, our method employs multimodality data into a single model, which reduces the cost of adding new SR objectives such as detection and classification tasks. The results reveal that the proposed method outperforms existing methods in objective and subjective evaluation as shown in Figure. \ref{Fig1}}.
\end{itemize}

The rest of the paper is drawn as follows. Section II
provides an overview of the related works. The proposed
The multimodal-Boost approach is detailed in Section III. Section IV contains the results and discussion. Finally, in Section V, the conclusion is given.

\begin{figure}[h]
\includegraphics[width=0.5\textwidth]{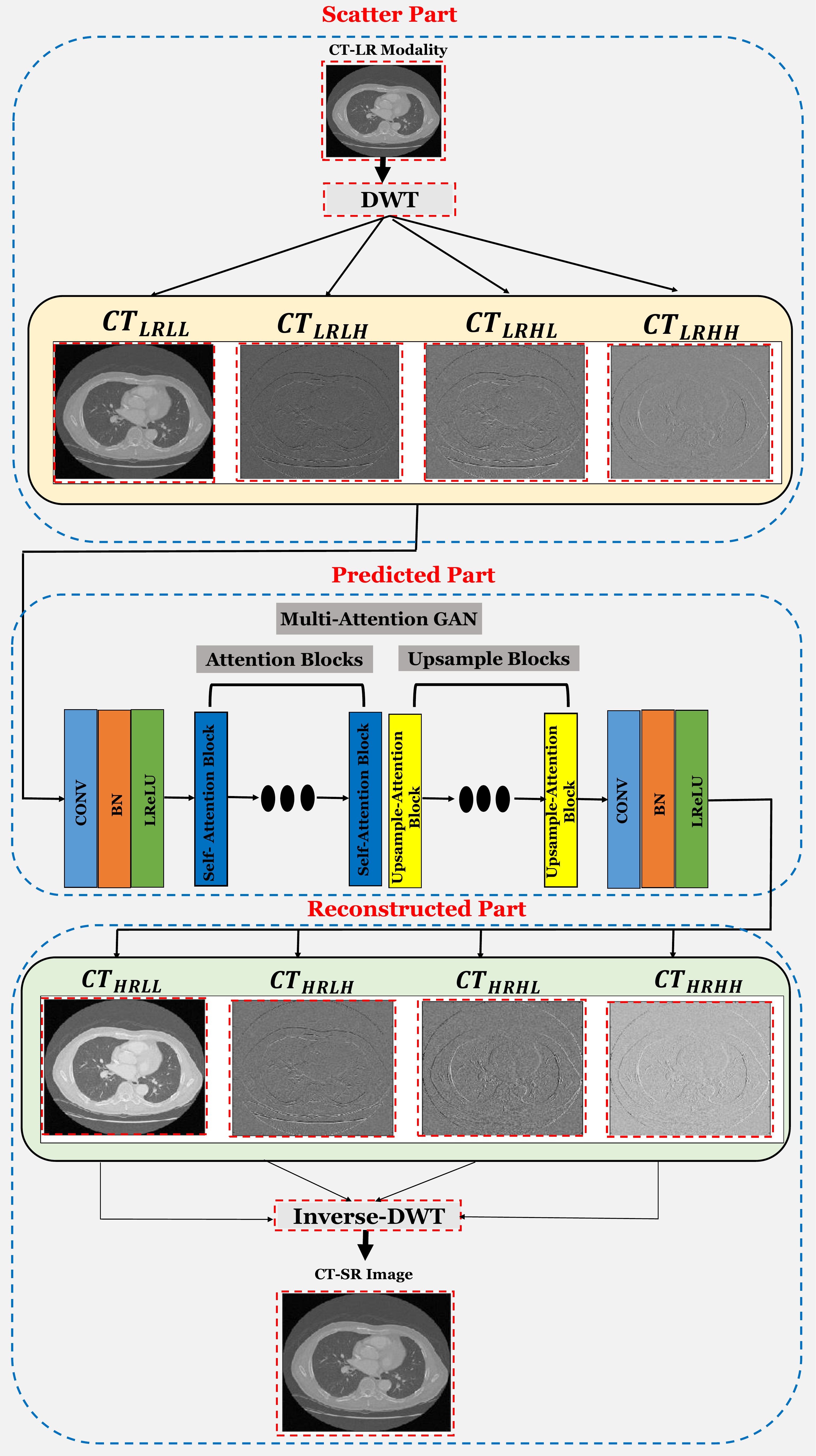}
\caption{Multimodal-Boost presents a three-part methodology flow chart: scatter, predicted, and reconstruct portions. The interpolated version of the LR image is broken down into four LR wavelet components in the dispersed section. In the predicting section, a developed Mult-Attention GAN SR network is used to estimate the HR wavelet component from the LR counterpart. The third portion is the reconstructed component, which uses 2DIWT to generate super-resolution images. The two forms of attention blocks are multi-attention and upsample-attention.}
\centering\label{Fig2}
\end{figure}

\section{Literature Work}
SR is undoubtedly a classical low-level vision problem which has been highly studied with emerging deep learning based solutions for a variety of applications. The problem of SR can be categorised into either single image SR or multiple image SR, our focus in this paper is on single image SR. Thus, in this section we review the state-of-the-art SR methods for both general and medical imaging application. 

In the domain of DL based SR, SRCNN by Dong et al\cite{dong2014learning} was the very first technique incorporating deep CNNs for the SR task. Following that several works were proposed with a typical pipeline of feature extraction, non-linear mapping and HR image reconstruction. However, such deeper level and complex architectures failed to extract the important multi-level features of a given LR image. Liang et al\cite{liang2016incorporating} identified the importance of edge based features for enhancing the SR task using sobel edge priors of LR input for training a deep SR model. Later, VDSR\cite{kim2016accurate}, EDSR\cite{lim2017enhanced}, and RFASR\cite{ahn2018fast} were proposed with the use of very deep networks involving residual skip connections and residual feature aggregation techniques to enhance the SR output. However, these methods did not yield admissible SSIM and PSNR due to deterioration of visual perception which resulted in over-smoothed SR output. Ledig et al\cite{ledig2017photo} improved the visual perception in SR problem by proposing a GAN based network namely SRGAN by combining adversarial and content loss of generated HR output. Moreover, Meta-SR\cite{hu2019meta} and DBPN\cite{haris2018deep} achieved improved results where the former proposed a solution for high magnification and arbitrary magnification SR and the later used a deep back-projection network to generate clear HR output. 

While all these methods work well on natural datasets, their performance is not admissible for medical imaging tasks. The methods on pixel space optimization\cite{dong2014learning,lim2017enhanced,ahn2018fast} generate a smooth output which lacks the important information, while the methods on feature space optimization\cite{ledig2017photo} could not generate realistic enough outputs. For this reason, these methods are not well suited for a critical problem like medical imaging SR. Although there exist several studies to tackle medical imaging SR with the techniques such as cycle GAN\cite{liu2021perception}, U-Nets\cite{park2018computed}, 3D convolutions\cite{chen2018efficient} and attention based networks\cite{he2020super}, none of them achieved acceptable clinical ability. Moreover, most of these works are suitable only for single modality and their performance degrades when doing inference on new modality dataset. Therefore, in this paper we aim to address the shortcomings of above works by utilizing the inherent capabilities of discrete wavelet analysis on embedding the LR input image which is then provided to multi-attention and upsample blocks of proposed GAN network to produce the corresponding HR output. It can be seen from results that our model efficiently utilizes the transfer learning technique to produce clear HR output on multiple medical imaging modalities.

\section{METHODOLOGY}
\subsection{SR Based on the DWT}
For low-level vision tasks like image denoising and super-resolution, choosing the proper wavelet can be difficult. Medical diagnostics, in particular, demands HR images with improved contextual features to provide better patient care. The WT stores detailed information about an image in many orientations to make the most of it. At each level of decomposition, the 2D-DWT produces four subbands that correspond to distinct frequency components and are referred to as approximate subbands, horizontal, vertical, and diagonal, containing complete edges information. Each subband contains image data with distinct characteristics that are important enough to offer precise image features. 

In practice, DWT is being used to send the input signal via the low-pass L(e) and high-pass filters H(e). The input signal's approximation and accuracy are then reduced by 2. $L(e)$  and $H(e)$, are the Haar wavelets' definitions:
\begin{equation}
  L(e)=
    	\begin{cases}
    	    1, & e=0,1 \\
            0,  &  otherwise\\
        \end{cases}
          H(e)=  
         \begin{cases}
    	 1,  &   e= 0 \\
        -1,  &   e= 0\\
         0,  &   otherwise\\
        \end{cases}
\end{equation}

\begin{figure*}[ht]
\includegraphics[width=1\textwidth]{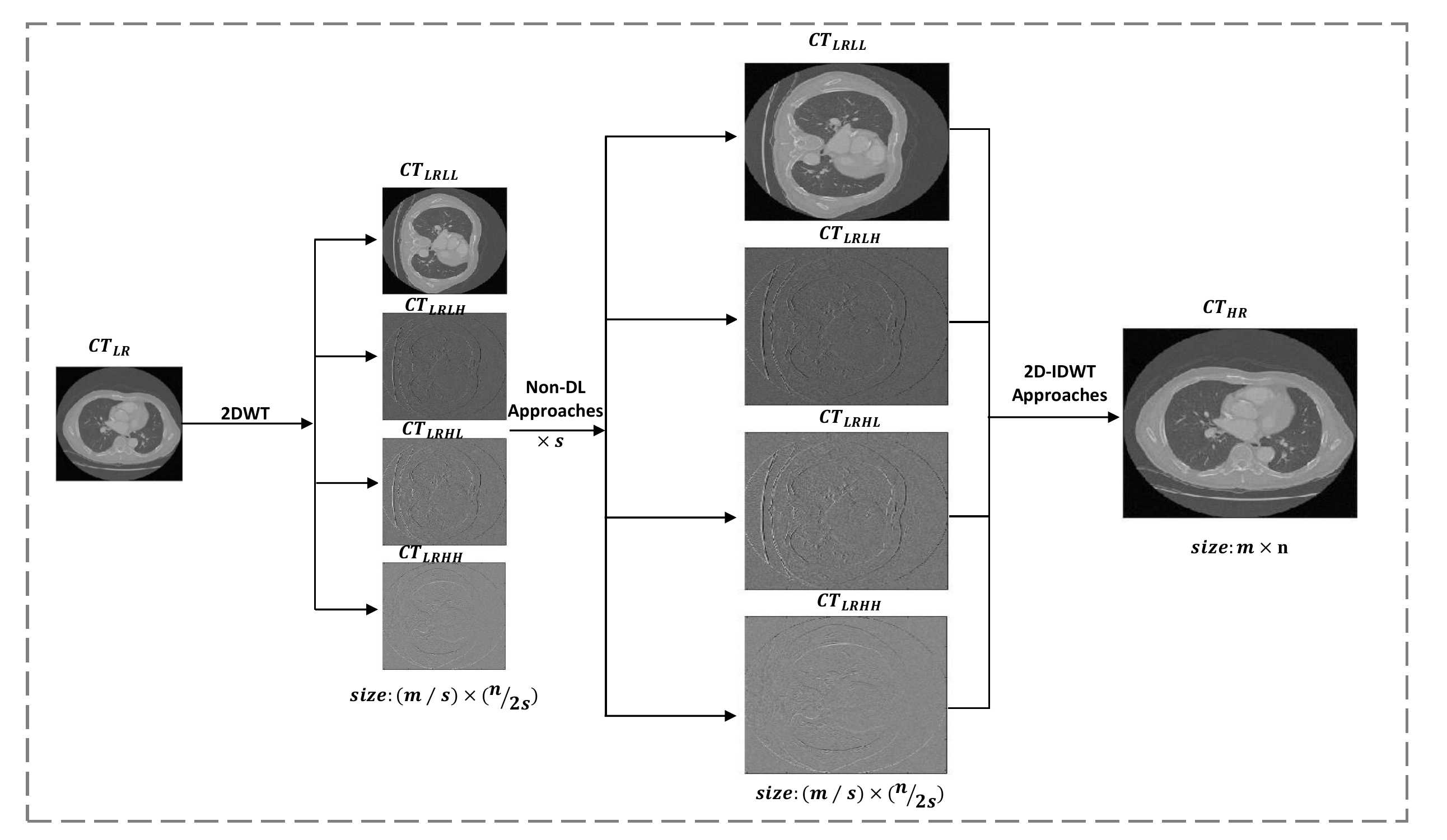}
\caption{level-1 decomposition of CT image. 2DWT extracts four subbands ($\textbf{CT}_{LL}$, $\textbf{CT}_{LH}$, $\textbf{CT}_{HL}$, and $\textbf{CT}_{HH}$) from the input CT modality image, which are subsequently fed to the Predicted Network for training.}
\centering\label{Fig3}
\end{figure*}

Generally, an image I(x,y), where xth and yth pixel values are in columns. The average components of the 2D-DWT are represented by four subbands: approximation, vertical, horizontal, diagonal, \textbf{LL},\textbf{LH}, \textbf{HL}, \textbf{HH}.

When we use 1-level 2D-DWT to an LL input image to predict the HL, LH, and HH subbands, we get the miss-ing information features of the LL image, as shown in the figure. To obtain SR results, we employ 2D-IDWT to acquire the image missing features information.
The coefficients of 2D-IDWT can be computed as below from the Haar wavelet.

\begin{equation}
    \begin{cases}
    	    A=a+b+c+d  \\
            B=a-b+c-d   \\
            C=a+b-c-d    \\
            B=a-b-c+d 
        \end{cases}
\end{equation}

where A, B, C, D, and a, b, c, d are the subbands pixel values. Interpolation-based SR approaches are limited in their capacity to reconstruct information appropriately. The result is not accurate because high-frequency domain information cannot be adequately recovered during SR. To improve super-resolved image performance, edges must be retained. The DWT was employed to keep the image edges features and texture details in high-frequency; the 2-DWT decomposed the $I_L$ image into the \textbf{LL}, \textbf{LH}, \textbf{HL}, \textbf{HH} subbands.
 Conventional DWT-based SR approaches combine DWT with Non-DL SR methods\cite{dharejo2021twist} . Let $I_G$ denote the image of the ground-truth HR in  $m \times n $  by, and $I_L$ symbolize LR in size $m \times n$ by, and $I_L$ by the same scale factor (s). Authorize ${I_{LR}}$ to represent $I_L$ up-scale ability, with $m \times n $ size, and  to display the \textbf {HR} image reconstructed, with m×n  size, via bicubic interpolation.

\begin{enumerate}
  \item  The four different subbands are achived form LR input image, as shown in Figure. \ref{Fig2}., i.e.,$\textbf{CT}_{LL}$, $\textbf{CT}_{LH}$, $\textbf{CT}_{HL}$, $\textbf{CT}_{HH}$  $\textbf{CT}_{LR}$ 

  \item  Finally using a discrete inverse WT, the HR image of $\textbf{CT}_{HR}$ is reconstructed (IDWT).

\end{enumerate}

\begin{figure*}[ht]
\includegraphics[width=1 \textwidth]{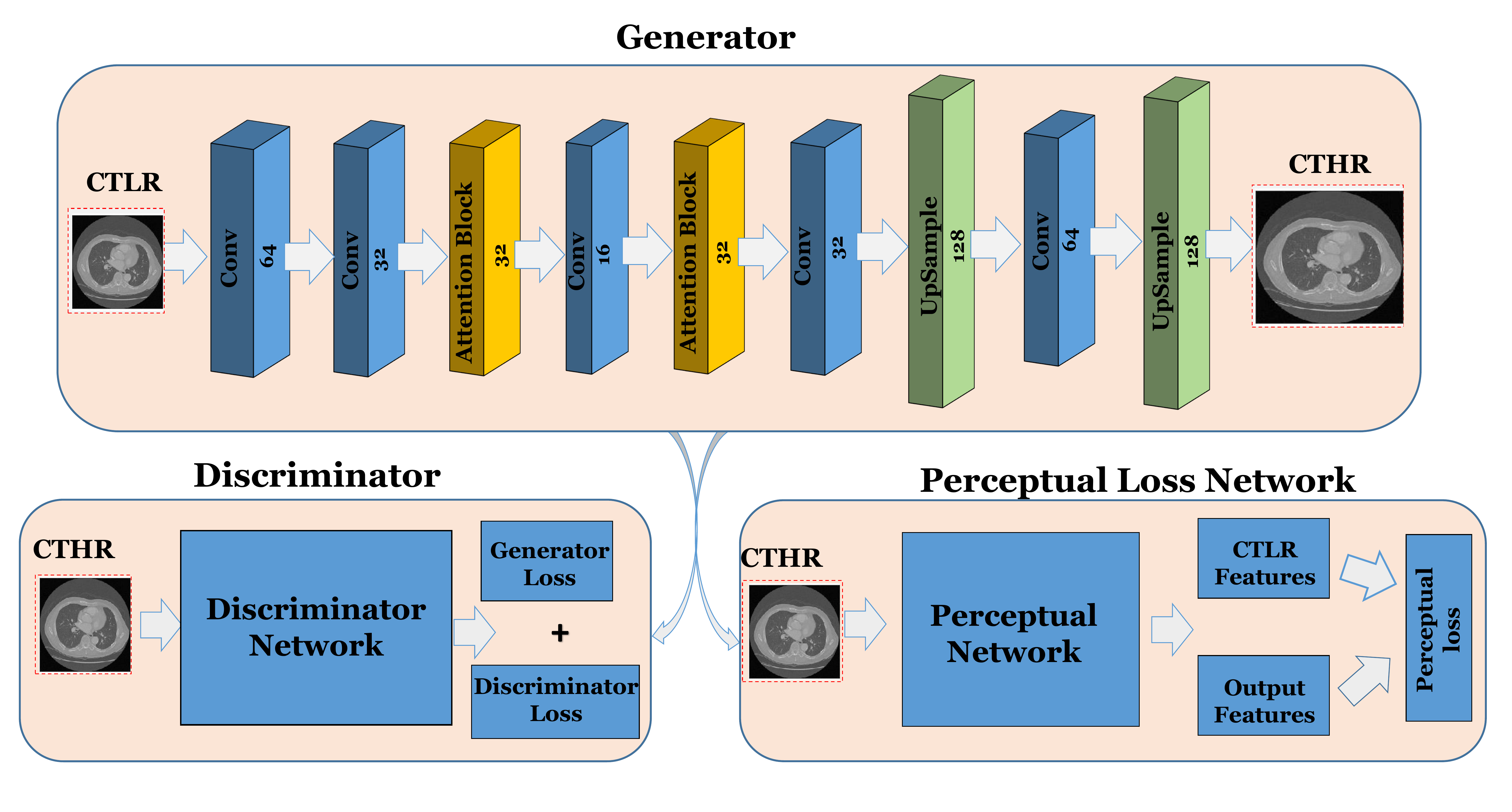}
\caption{Proposed GAN network with multiple attention and upsample blocks.}
\centering\label{Fig4}
\end{figure*}

\subsection{Multimodal Muti-Attention GAN}
A generator and a discriminator are the two fundamental components of a GAN. Samples z from a preceding noise distribution $\textbf{p}_{noise}$ are sent into the generator G. The generator is performed with any model distribution $\textbf{p}_{model}$ in the data space $\hat{x}=G(z)$. The discriminator is a binary classifier whose function is to distinguish between real and fake data on generated samples $x\sim P_{data}, D(\hat {x
})=1$ and $x\sim P_{data}, D(\hat{x})=0$.
Convolutional layers are featured in almost all GAN-based image restoration models \cite{salimans2016improved}\cite{creswell2018generative}. GANs evaluate information in a local neighborhood while deploying convolutional layers is computationally expensive for long-range modeling relationships in images. Our proposed approach uses DWT to generate wavelet subbands to improve spatial resolution (super-resolution). Self-attention is introduced in to the GAN framework, allowing the generator and the discriminator to efficiently describe interactions between widely separated spatial regions, as shown in Figure. \ref{Fig4}.

To calculate the attention blocks, the image features from previously hidden layers  $x\in \mathbf{R} ^{C\times N} $ are transferred to feature spaces \textbf{f;g}, where $\textbf{f(x)}=W_{fx};g(x)=W_g$ 
\begin{equation}
\beta_{i,j}= \frac{\exp(S_{i,j})} {\sum_{i=1}^N \exp(S_{i,j})}  = {f(x_i)^T} {g(x_i)}
\end{equation}

When constructing the $\textbf{jth}$ area, $\beta_{i,j}$ shows how much attention the model pays to the $\textbf{ith}$  place. Where C is Considering N is the number of locations of features from the previously hidden layer, the output of the attention layer can be expressed as, $o$=$(o_1,o_2,…,o_j,…,o_N )$ $\in \mathbf{R}^{C \times N}$ where,

\begin{equation}
O_j= \mathcal {V}(\sum _{i=1} ^N \beta_{i,j} h(x_i)), h(x_i)=W_h X_i, \mathcal {V}=W_h X_i
\end{equation}

The learned weight matrices $\textbf{W}_g$  $\in \mathbf{R}^ {L \times C}$, $\textbf{W}_f$  $\in \mathbf{R}^ {L \times C}$, $\textbf{W}_h$  $\in \mathbf{R}^ {L \times C}$, and $\textbf{W}_v$ $\in \mathbf{R}^ {L \times C}$  are implemented as 1×1 convolutions.  We used varied channel numbers of L=C/K, in all of our experiments, where k=1,2,4,8 thus after a few ImageNet training epochs, we chose $k=8 (i,e.,L= \frac{C}{8})$  and found no significant performance loss. We also add back the input feature map after multiplying the output of the attention layer by a scale parameter. As a result, the final outcome is as follows:

\begin{equation}
Y_i= \alpha o_i + x_i 
\end{equation}

\begin{figure*}[ht]
\includegraphics[width=1 \textwidth]{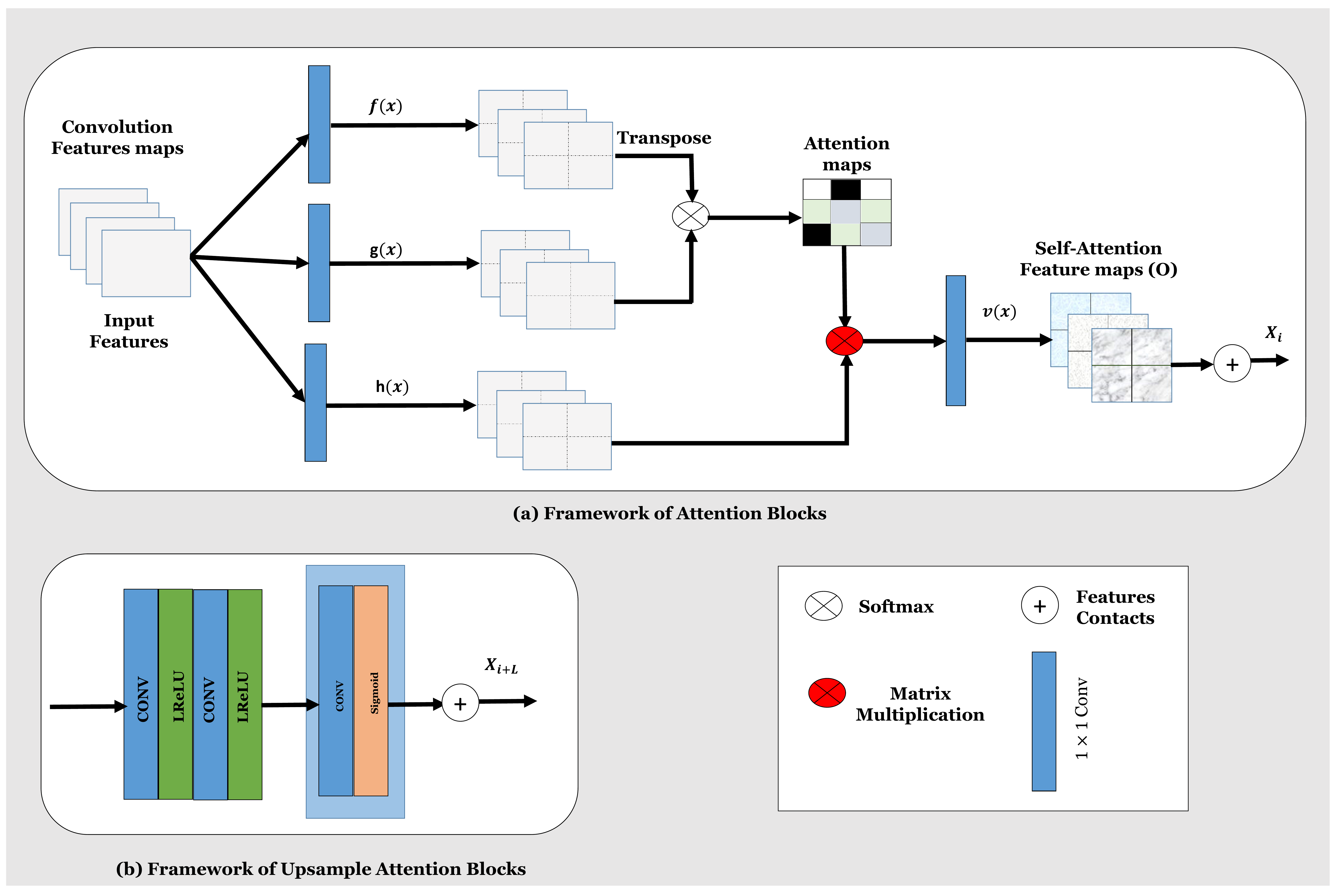}
\caption{Mechanism of attention blocks, whose goal is to boost network performance by improving super-resolution, where (a) displays the mechanism of Self-attention blocks and (b) illustrates the structure of Upsample blocks.}
\centering\label{Fig5}
\end{figure*}

Where $\alpha$ is a scalar parameter initially set to 0, allowing the network to depend on local inputs first and then progressively learn to give non-local data to more weights. The attention block used in our method is shown in Figure \ref{Fig5}(a). Upsampling attention blocks, which tend to improve the resolution for CT images on wavelet subbands input features, are another attention block employed in our predicted part. The spatial resolution of the feature map was further improved by performing local neighborhood interpolation on the wavelet subband input features. These are fed to the leaky convolution ReLU, and the final feature map $(f_{Upsample}$ $\in \mathbf{R}^ {L \times C}$ Upsample attention block can be expressed as;

\begin{equation}
F_{map}= Sigmoid(Conv(1\times 1))\times f_{upsample}
\end{equation}

$F_{map}$ is referred to output of  pixel attention module and  $\textbf{(Conv(1×1))}$ represent the convolution operation with kernel size 1×1 so we can say the $F_{map}$ $\in \mathbf{R}^ {W \times L \times 1}$. The values of the feature map fall between 0 and 1. The feature map can be correctly changed at the component level to achieve superior SR results. The Upsample attention block is given in Figure \ref{Fig5}(b). 
\begin{figure*}[ht]
\includegraphics[width=1 \textwidth]{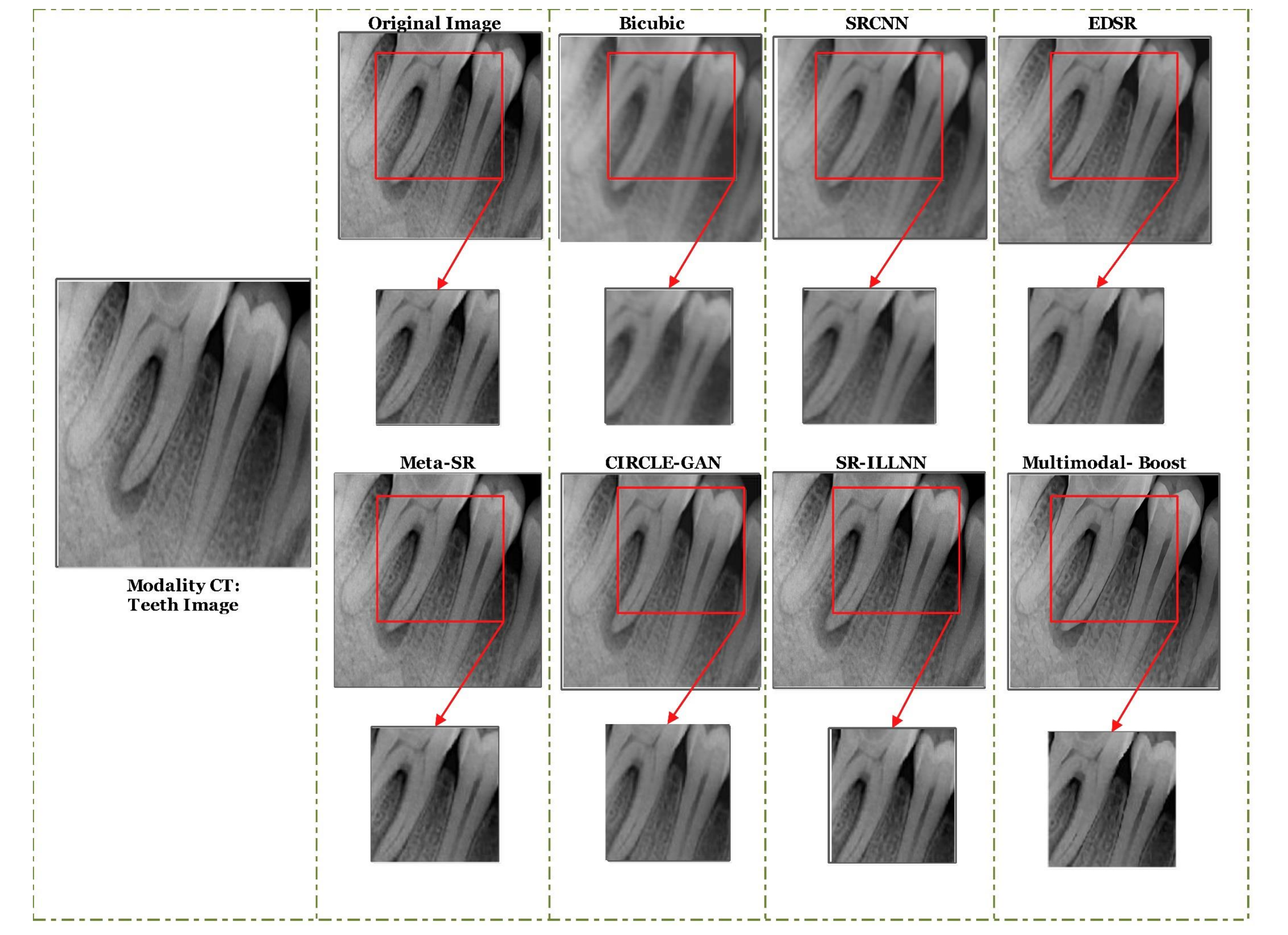}
\caption{Visual results of our proposed method (Multmodal-Boost) compared to our State-of-art methods. We  enlarge the particular region to notice the differences between the outcomes more clearly on CT teeth.}
\centering\label{Fig6}
\end{figure*}

\subsection{The Reconstruction Part}
To reconstruct the HR super-resolved image $CT_{HR}$, we use a 2D-IDWT inverse wavelet onto four components $\boldsymbol{CT_{LL}}, \boldsymbol{CT_{LH}}, \boldsymbol{CT_{HL}}$, and $\boldsymbol{CT_{HH}}.$

\subsection{VGG based Perceptual Loss }
\begin{figure}[h]
\includegraphics[width=0.5 \textwidth]{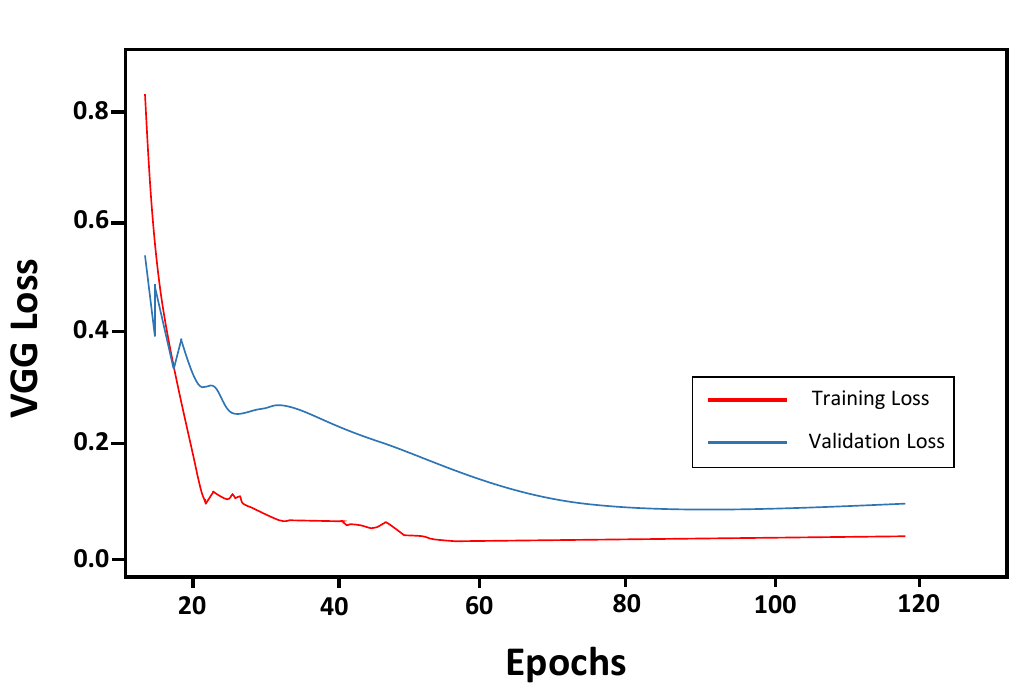}
\caption{VGG-16 Loss curves obtained on natural images.}
\centering\label{Fig7}
\end{figure}

Perceptual loss is one of the best metrics for measuring image similarity when CNN algorithms are applied to the input image. It is utilized in various applications such as image denoising, image dehazing, image translation, and image super-resolution. Compared to  Mean Squared Error (MSE) loss \cite{johnson2016perceptual}, the perceptual loss is more robust to several problems such as over-flattening and distortion \cite{shan20183}\cite{chen2017face}; hence for image SR, the perceptual loss is more powerful to search spatial resolution similarities between two images. Many CNN networks employ VGG-loss to measure perceptual loss; however, VGG-11, VGG-16, and VGG-19 are pre-trained networks that achieved amazing results using natural image datasets \cite{he2018amc}. VGG-loss is a term that can be stated as follows:

\begin{equation}
L_{VGG}= \mathbb{E}_{(I_{LR, HR})} [\frac{\|VGG(g(I_{LR}))-VGG(I_{HR})\|_F^2}{DHW}]
\end{equation}
       
where D; H; W denotes the computed tomography (CT) image depth, height, and width. VGG was trained for image classification using natural mages; therefore, it generates features not relevant to CT image super-resolution. This is one of the possible pitfalls with VGG-loss \cite{jo2020investigating}. Due to the scarcity of labeled CT images, VGG for CT datasets is a challenging task. We proposed a trained perceptual network to handle this challenge to extract a compressed encoding from CT data input and reconstruct an image comparable to the original image. In our case we used VGG-16 in the described perceptual loss network as shown in Figure \ref{Fig4}. The perceptual network comprises six convolution layers, each with 32, 32, 64, 64, 128, 128 filters. There is also a max pooling layer with a kernel size of 2 and a stride of 2. The perceptual loss convolution layers were utilized, followed by the ReLU activation function. In this design, we used a 33 filter size and a stride of 1.  To extract the features, we trained the perceptual net-work to determine the perceptual loss. Figure \ref{Fig7} shows the perceptual loss based on VVG-16.
\begin{equation}
L_{prec}= \mathbb{E}_{(I_{LR, HR})} [\frac{\|\gamma (g(I_{LR}))-\gamma (I_{HR})\|_F^2}{DHW}]
\end{equation}

Where $\gamma$ is the pre-trained encoder network. 

\begin{figure}[h]
\includegraphics[width=0.5 \textwidth]{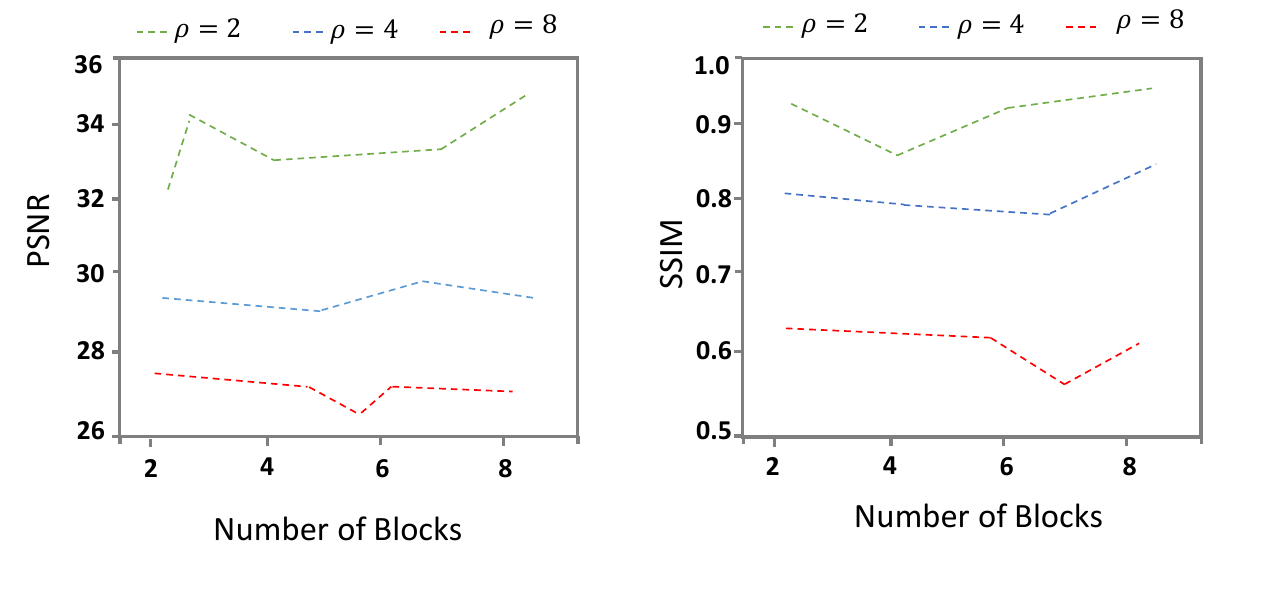}
\caption{Performance evaluation of our model with different numbers of multi-attention blocks.}
\centering\label{Fig8}
\end{figure}

\subsection{Model Optimization}
The proposed multimodal multi-attention model loss is equal to the sum of the perceptual and generator-discriminator losses, as shown below:

\begin{equation}
\min_{g}\max_{d}L_{WGAN}(g,d)+\beta L_{perc}(g)
\end{equation}
$L_{WGAN} (g,d)$ is a Wasserstein GAN \cite{arjovsky2017wasserstein}\cite{isola2017image} optimizer, the weighted parameters are represented by the $\beta$, which claims to stand for substitution between WGAN-Loss and perceptual loss, and the letters g and d, which stand for generator and discriminator, respectively. $L_{perc} (g)$ is the perceptual loss. $L_{WGAN} (g,d)$ is in addition to the usual Wasserstein distance and the regularization gradient penalty, which is represented as,

\begin{align*}
\min_{g}\max_{d}L_{WGAN}(g,d)= (- \mathbb{E}_{I_{HR}}[d(I_{HR})]+ \\\mathbb{E}_{I_{LR}}[d(g(I_{LR})])+
\lambda \mathbb{E}[(\|\bigtriangledown_{\hat{I}}d(\hat{I})\|_2-1)^2]
\end{align*}

Where $\mathbb{E}_a [b]$ is the expression of b as a function of a, $\lambda$ represents the weighted parameter, $\hat{I}$ denotes the generated and real images in uniform sampling from a range of [0,1], and $\bigtriangledown$ is the gradient.

\begin{figure*}[ht]
\includegraphics[width=1 \textwidth]{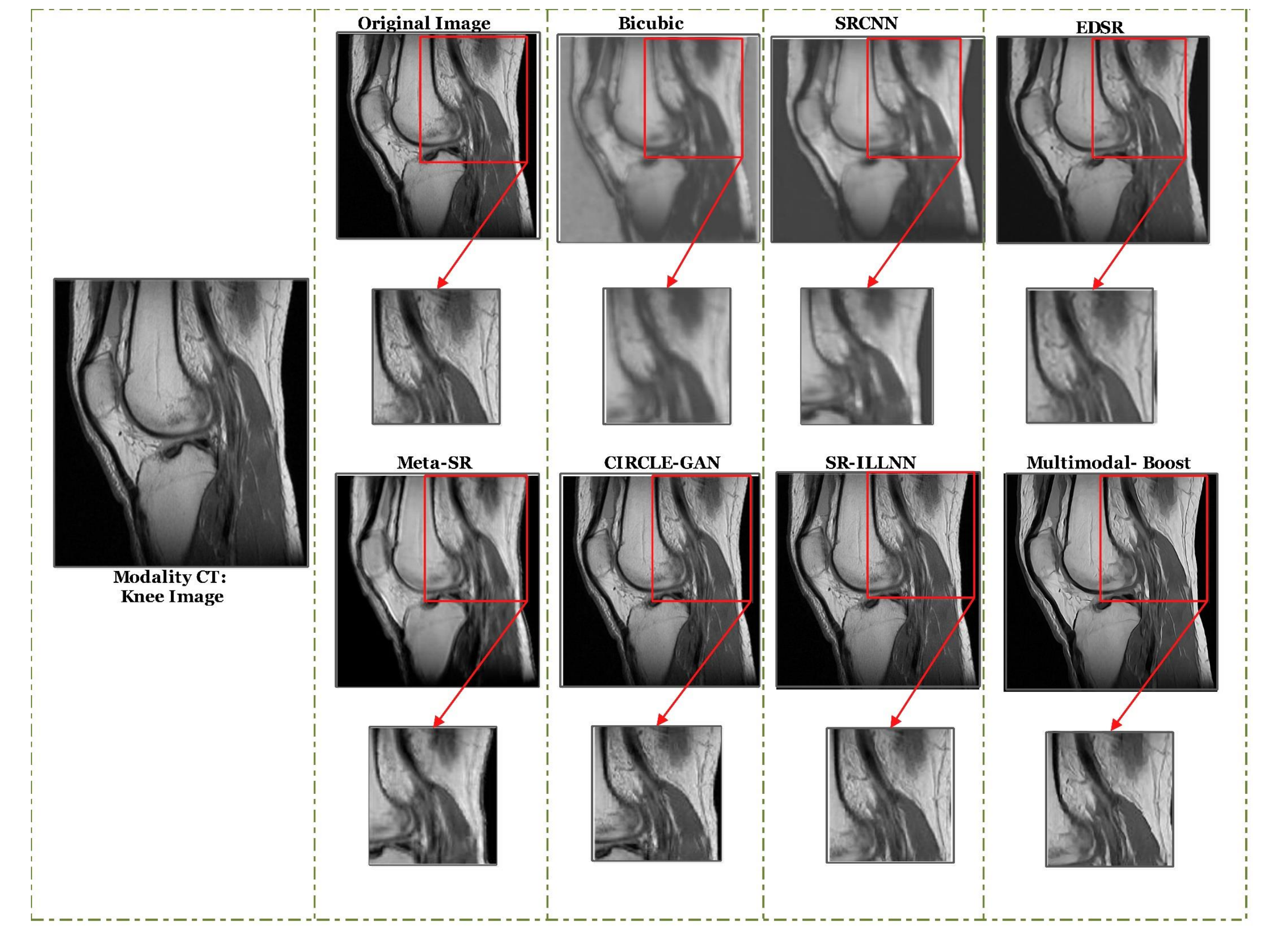}
\caption{Visual results of our proposed method (Multimodal-Boost) compared to our State-of-art methods. We  enlarge the particular region to notice the differences between the outcomes more clearly on CT knee image.}
\centering\label{Fig9}
\end{figure*}

\begin{figure*}[ht]
\includegraphics[width=1 \textwidth]{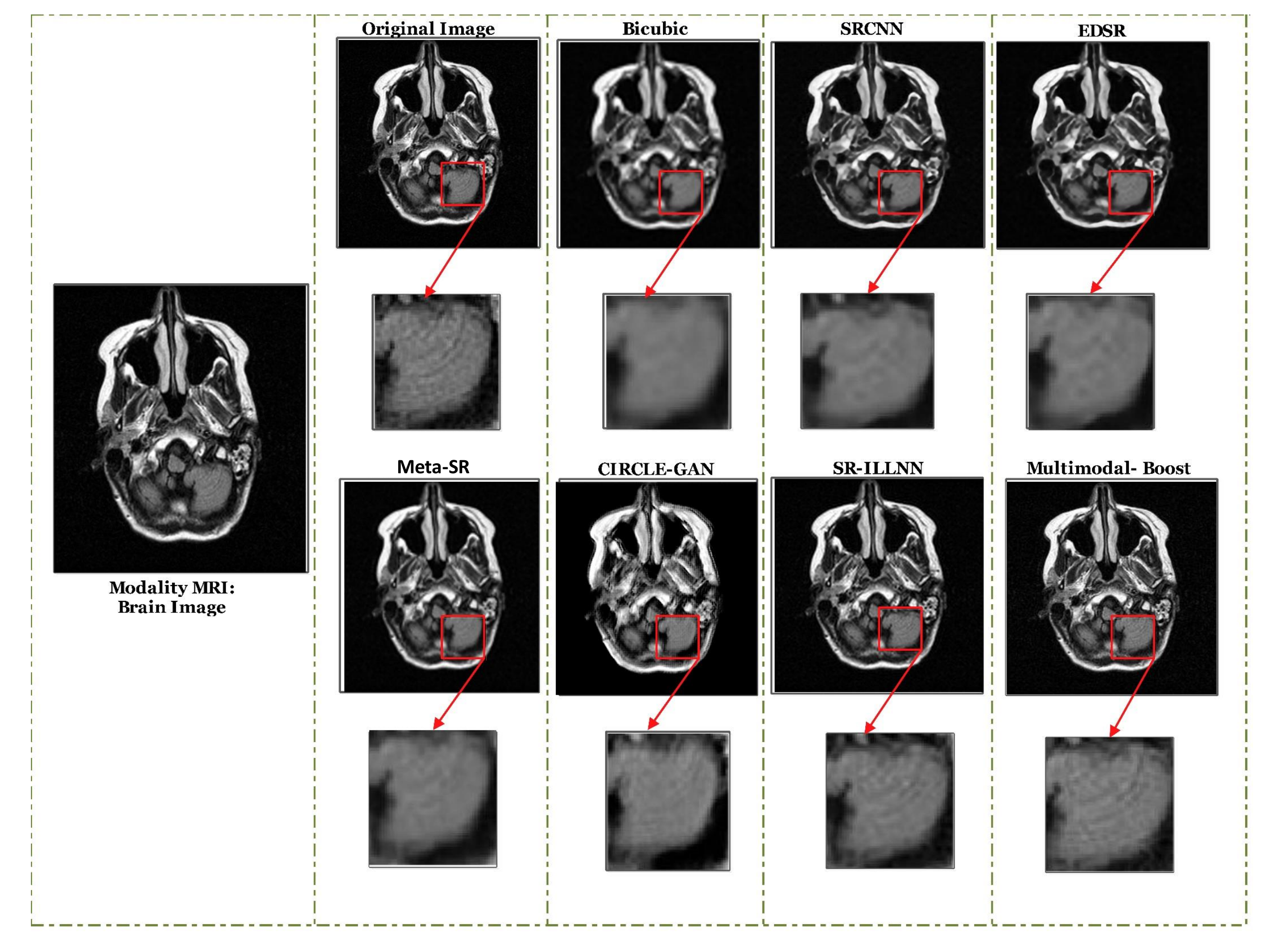}
\caption{Visual results of our proposed method (Multmodal-Boost) compared to our State-of-art methods. We  enlarge the particular region to notice the differences between the outcomes more clearly on MRI-Brain image.}
\centering\label{Fig10}
\end{figure*}

\section{RESULTS AND DISCUSSIONS}
We introduced above a Multimodal-Boost framework for super-resolution tasks in medical images. It comprises a new neural network for SR medical image reconstruction, pair-wise attention blocks, and a novel GAN-based loss function. The proposed model is efficient with texture details and realistic for viewing reconstructed images to a large extent. Since super-resolution is an inverse problem with unpredictable solution \cite{lim2017enhanced} , the output of the SR image contains substantially more information than the matching LR image. Each LR feature from wavelet subbands was regarded as a training sample during Multimodal-Boost model training and fed to the training phase. Loss functions are employed to ensure generated SR images as close to HR. Since working with LR images for diagnosis is too challenging, SR images are highly rated for medical applications, and they include enough information for radiologists to make precise conclusions. The information was stored in hidden layers by SR DNN models, which resulted in high frequency or HR images with superior texture and edge details. Medical images are too complicated to handle compared to natural images, making machine learning models relatively challenging. The higher magnification area in Figure \ref{Fig6}, Figure \ref{Fig9}, and Figure \ref{Fig10} shows the differences between each approach. At the same time, Meta-SR achieves excellent performance than other techniques; it still struggles to generate pretty small textures, whereas our proposed method takes advantage of wavelet transform and obtains more excellent texture details. 
We also perform visual comparisons on other images using our Multimodal-Boost and other SR techniques, as seen in Figure "teeth-image" Figure \ref{Fig6}. All previous SR algorithms produce blurry outputs and fail to recover detailed information in the magnified image, while Meta-SR and our new approach Multimodal-Boost can only reconstruct a clean image with sharp lines. We evaluate the SR results of the image MRI Modality to support the multimodality and discover that SRCNN, EDSR, and Meta-SR all fail to recover the clear edges. It is indeed worth remembering that the texturing details they acquired are incorrect. CIRCLE-GAN and SR-ILLNN, on the other hand, can rebuild more compelling results that are consistent with ground truth, but they fail to provide more precise edge information. On the other hand, our technique does better when it comes to restoring edges and texture details.

\subsection{Training Details}
When training SR models with pairs of LR and HR images, downsampling is typically utilized. On the other hand, medical datasets have fixed in-plane resolutions of roughly 1mm and lack ultra-high spatial resolutions. As a result, transfer learning is one of the methods for effectively training models using external datasets, which has proven to be very useful in remote sensing and medical applications. We used massive datasets DIV2K \cite{timofte2017ntire} with $2000 \times 1400 $ pixels of good quality to train our proposed technique to use transfer learning. We randomly cropped $56\times 56$ sub-pictures from DIV2K training images for training. Preliminary, we obtain the trained and optimized network by pre-training the proposed model in a, particularly defined manner. The pre-trained network is fine-tuned by selecting the Shenzhen Hospital datasets \cite{jaeger2014two}, containing 662 X-ray images used for training and testing. All images were scaled to $512 \times 512$ pixels, and three modalities, including Montgomery County X-ray, Teeth,, and knee images, were chosen to test our model. MRI brain scans from the Calgary Campinas repository were another dataset we used to evaluate our proposed approach. This modality is produced using a 12-channel head-neck coil on an MR scanner (Discovery MR750; General Electric Healthcare, Waukesha, WI). All experiments have been carried out on a Windows-based machine with an Intel(R) Core(TM) i5-7300HQ CPU running at 3.40GHz and an NVIDIA GeForce GTX 1080-Ti graphics card. The setup also makes use of MATLAB 2019 with CUDA Toolkit and Anaconda.

\begin{table*}[ht] 
\caption{DETAILS OF ALL TRAINED MODELS AND DIFFERENT LOSS MEASURES} 
\centering 
\begin{tabular}{|c| c | c | c |c |  } 
\hline  Experiments & Generator Network & Loss Function & WGAN\\ 
\hline  CNN-VGG & CNN & $L_{VGG}$ & no\\
\hline  WGAN & CNN & $L_{WGAN}$ & yes\\ 
\hline  Perceptual  & Multi-attention & $L_{perc}$ & no\\ 
\hline  WAGN-VGG & CNN & $L_{WGAN}$ + $L_{VGG}$ & yes\\ 
\hline  WAGN-MA-P & Multi-attention CNN & $L_{WGAN}$ + $L_{perc}$ & yes\\ 
\hline  
\end{tabular} 
\label{tab:var} 
\end{table*} 

\begin{table*}[ht] 
\caption{NETWORK COMPLEXITY AND INFERENCE SPEED ON SHENZHEN HOSPITAL DATASETS. MULTIMODAL-BOOST, IN COMPARI-SON TO SRDESNSENET AND ESRGAN, TAKES MORE TRAINING TIME WHILE TAKING LESS TIME THAN EDSR AND CIRCLE-GAN WITH TRANSFER LEARNING. FURTHERMORE, UNLIKE EDSR AND META-SR, WHICH ONLY WORK WITH A SINGLE MODALITY, IT USES MULTI-MODALITY DATA IN A SINGLE MODEL, LOWERING THE COST OF ADDING NEW SR TASKS} 
\centering 
\begin{tabular}{|c| c | c | c |c |  } 
\hline  Methods & Number of parameters (M) & Memory size (MB) & Inference time(s) & Number of $log_{10}$ flops\\ 
\hline  SRCNN \cite{dong2015image} & 0.059 & 13.68  & 0.0113 & 17.4\\
\hline  EDSR \cite{lim2017enhanced}  & 32.55 & 266.78  & 2.178 & 24.2\\ 
\hline  Meta-SR \cite{hu2019meta} & 4.075 & 32.64 & 1.0156 & 25.1 \\ 
\hline  GAN-CIRCLE \cite{you2019ct} & 55.88 & 457.88 & 3.0172 & 24.8\\ 
\hline SR-ILLNN \cite{kim2021single} & 0.441 & 16.44 & 0.0123 & 28.1\\ 
\hline  Multimodal-Boost &	21.68 &	105.68 & 2.0189 & 28.3\\ 
\hline  
\end{tabular} 
\label{tab:var} 
\end{table*}

\begin{table*}[ht] 
\caption{OUR PROPOSED METHOD OUTPERFORMS SRCNN, EDSR, META-SR, CIRCLE-GAN, SR-ILLNN AND BICUBIC INTERPO-LATION ON MULTI-MODAL IMAGES SUCH AS TEETH, KNEE, CT SCAN IMAGES, AND CARDIAC MR BRAIN SCANS IMAGE IN TERMS OF BOTH QUALITATIVE AND QUANTITATIVE OUTCOMES. A GREATER PSNR INDICATES BETTER SR IMAGE RECON-STRUCTION QUALITY, WHILE A HIGHER SSIM INDICATES BETTER PERCEPTUAL QUALITY. BOLD REPRESENTS THE BEST PERFORMANCE.}

\centering 
\begin{tabular}{c| rrrrrr } 
\hline 
 &\multicolumn{2}{c}{CT: Modality
(Teeth-Image)
} & \multicolumn{2}{c}{CT: Modality
(Knee-Image)
} & \multicolumn{2}{c}{MRI: Modality
(Brain-Image)
}\\  
State-of-art Methods 
 & PSNR & SSIM &PSNR & SSIM & PSNR & SSIM \\ [0.25ex] 
\hline   
Bcubic  & 26.30 & 0.825 & 25.10 & 0.815 & 25.22 & 0.842\\  
SRCNN \cite{dong2015image} & 27.45 & 0.845 & 26.69 & 0.853 & 26.99 & 0.862\\ 
EDSR \cite{lim2017enhanced} & 29.54 & 0.873 & 28.52 & 0.868 & 28.72 & 0.890\\  
Meta-SR \cite{hu2019meta}  & 31.55 & 0.919 & 32.22 & 0.913 & 33.52 & 0.912 \\  
CIRCLE-GAN \cite{you2019ct} & 32.44 & 0.902 & 33.33 & 0.921 & 33.93 & 0.929\\ 
SR-ILLNN \cite{kim2021single} & 34.57 & 0.921 & 35.76 & 0.925 & 35.41 & 0.931\\ 
\textbf {Multimodal-Boost} & \textbf{36.32} & \textbf {0.937} & \textbf{36.97} & \textbf {0.931} & \textbf {37.46} & \textbf {0.941}\\ 

\hline 
\end{tabular} 
\label{tab:hresult} 
\end{table*} 

\subsection{Transfer Learning}
Transfer learning is a technique that improves the performance of deep neural networks by utilizing knowledge learned from natural image data sets as initial training data. Due to the different distributions of two types of images, directly applying a trained model using natural datasets to medical images will not work, so transfer learning is a feasible solution. Transfer learning has various advantages such as:
\begin{itemize}
    \item The ability to borrow high-frequency information from natural image datasets, which boosts the proposed method's performance to reconstruct the HR image form LR image.
    \item It helps in faster model convergence.
    \item It improves the model accuracy.
\end{itemize}
To achieve high image resolution for diagnostic, our suggested Multimodal-Boost model leveraged transfer learning to integrate shared and supplementary information from diverse modalities. In order to assess each modality and the detailed information inside the image patches, we used 16 to 128 batch sizes for the HR patches in model training. When the epochs in both modalities, CT-Images from Shenzhen Hospital dataset \cite{jaeger2014two} and MRI brain scans from CalgaryCampinas, reach 180, the training process is finished. We train our model with ADAM optimizer \cite{kingma2014adam} by setting $\beta_1 = 0.9$, $\beta_2= 0.999 $, and $ \in =10^{-8}$. 
Our primary goal is to improve multimodal medical image quality. We trained the model with DIV2K before fine-tuning it with CT-image dataset, which contains 662 X-ray images. We also tested our model using MRI brain imaging so that the previous modality information can be used to reconstruct the next modality image. According to the proposed results, transfer learning knowledge considerably improved performance as shown in Figure \ref{Fig11} and Figure \ref{Fig12} . The DIV2K dataset \cite{timofte2017ntire} is being used for training since it is a recently proposed high-quality 2k resolutions image dataset for image enhancement. There are 800 training images, 100 validation image.
\begin{figure}[h]
\includegraphics[width=0.5 \textwidth]{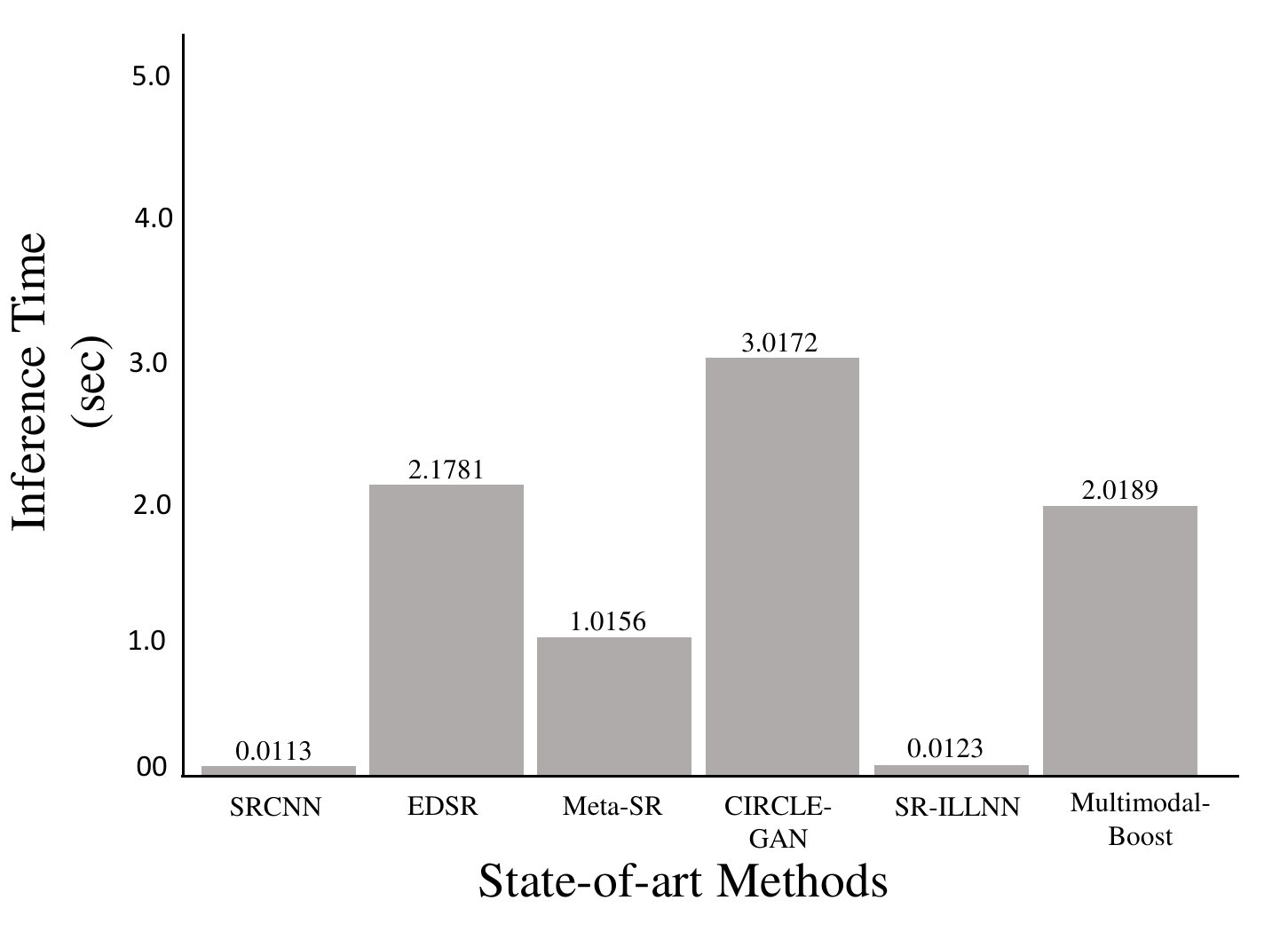}
\caption{Inference speed on Shenzhen Hospital dataset.}
\centering\label{Fig13}
\end{figure}
\subsection{Comparison with State-of-art Methods}
We analyzed and compared the proposed method with deep learning models including SRCNN \cite{dong2015image}, EDSR\cite{lim2017enhanced}, Meta-SR \cite{hu2019meta}, CIRCLE-GAN\cite{you2019ct}, SR-ILLNN \cite{kim2021single} and bicubic interpolation. These methods were designed for natural images DIV2K \cite{timofte2017ntire}, which have much bigger dimensions than the medical images we used.  We employed transfer learning and retrained the models with smaller chunks of medical image datasets to make them work more robustly. We used the same hardware and experiment conditions to make a fair comparison. We employed the MSE loss; however, it was not very reliable due to distortion and over-flattening. As a result, we proposed perceptual loss $L_{perc}(g)$ using VGG-16 training, and then we improved our network with total loss $L_{WGAN} (g,d)+\beta L_{perc} (g)$, which is the sum of perceptual loss and generative discriminator loss function. The proposed model (Multimodal-Boost) is compared against state-of-the-art approaches in Tables 2, and the visualization results are shown in Figure \ref{Fig6}, Figure \ref{Fig9}, and \ref{Fig10}.

\begin{figure}[h]
\includegraphics[width=0.5 \textwidth]{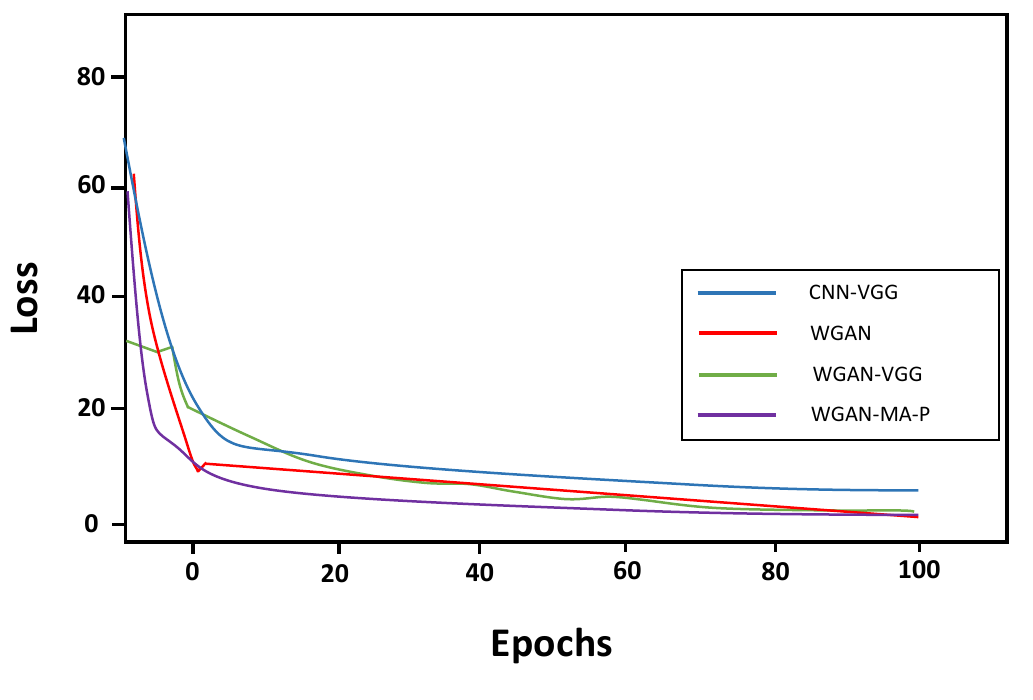}
\caption{Model loss over different experimental networks.}
\centering\label{Fig11}
\end{figure}
\begin{figure}[h]
\includegraphics[width=0.5 \textwidth]{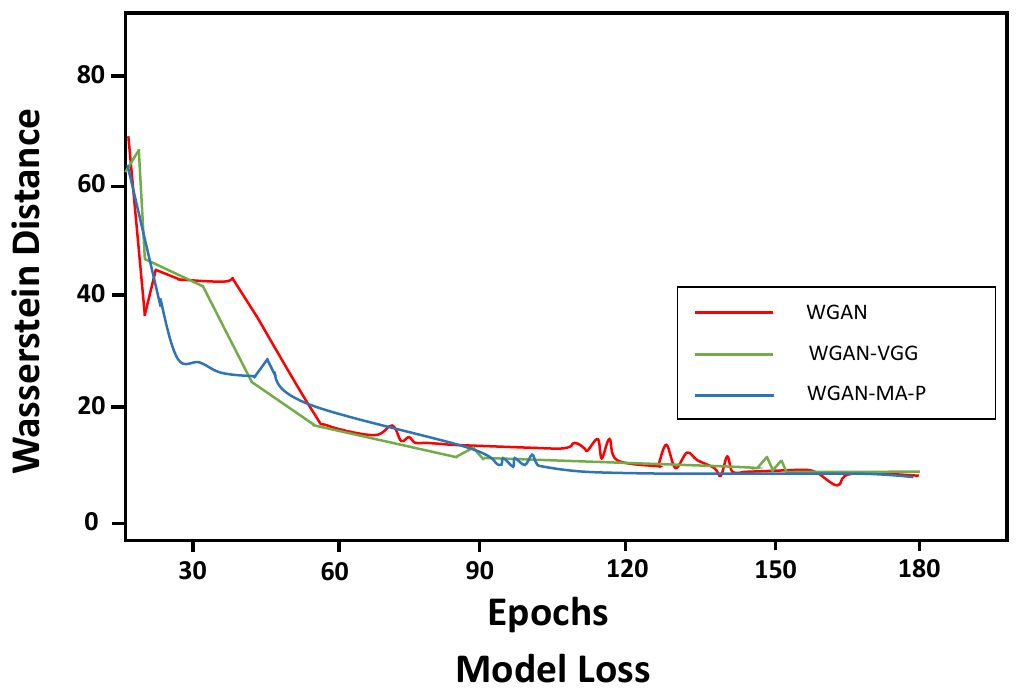}
\caption{Network convergence over different loss functions on Wasserstein Estimation.}
\centering\label{Fig12}
\end{figure}

For comparison, we conduct experiments on numerous networks with varying configurations, presented in Table I and Figure 
\ref{Fig11}. We witnessed that WGAN-based losses converge faster than non-WGAN-based losses. Finally, combining WGAN and perceptual loss results in a shorter distance and greater convergence, which we apply to all losses in model optimization, as shows in Figure \ref{Fig12}.

\subsection{Convergence}
The convergence curve of VGG-loss in WGAN-MA-P has the fastest convergence across all types of approaches, as shown in Figure \ref{Fig11}. WGAN-MA-P loss is quite similar to WGAN- VGG, although the latter achieves lower. However, a reduced VGG loss does not always imply improved performance. According to our findings, the VGG-loss network's spatial resolution diminishes its benefits while still causing some loss. Finally, we show how the WGAN-MA-P loss converges very quickly with shorter Wasserstein distances, resulting in state-of-the-art performance compared to other WGAN-based alternatives. The network complexity of methodologies is measured in terms of the number of parameters, memory, inference time, and number of flops as shown in Table 2. Compared to SRCNN\cite{dong2015image}, Meta-SR \cite{hu2019meta}, SR-ILLNN \cite{kim2021single}, Multimodal-Boost has many parameters; however, it has fewer parameters than EDSR \cite{lim2017enhanced} and CIRCLE-GAN  \cite{you2019ct}. Furthermore, on the same hardware, the proposed method takes 18\% longer to train than Meta-SR. As a result, further work on the architecture, such as model compression, should be done in the future.

\subsection{Quantitative Analysis}
We evaluated our approach on several attention blocks $(\rho=2, 4, 8)$ to investigate how well it performed. In terms of PSNR and SSIM, we found that the proposed Multimodal-Boost technique outperforms all others. When $\rho $ is increased, the proposed method performance degrades significantly, as illustrated in Figure \ref{Fig8}. Table 3 shows that the proposed strategy produces the maximum PSNR. For three images, SRCNN \cite{dong2015image} has the lowest PSNR, although it is better than Bilinear interpolation. The proposed method achieves 1.54–4.89 greater PSNR index when compared with recent methods EDSR\cite{lim2017enhanced}, Meta-SR\cite{hu2019meta}, CIRCLE-GAN\cite{you2019ct}, and SR-ILLNN \cite{kim2021single}. The quantitative performance metric is the peak signal-to-noise ratio (PSNR). The PSNR is defined as follows: Given a ground-truth image S and its reconstructed image $\hat{S}$, with $M\times N$ pixels size,

\begin{equation}
PSNR(S,\hat{S})= 10log_{10}\frac{225^2}{MSE(S,\hat{S)}}
\end{equation}

The Structural Similarity Index Measure (SSIM) is a popular tool for assessing the quality of high-resolution reconstructions. The SSIM index can be expressed mathematically as,

\begin{equation}
SSIM= \frac{(2\mu_x \mu_y+C_1)(2\sigma_{xy}+C_2)}{(\mu_x^2+\mu_y^2+C_1)(\sigma_x^2+\sigma_y^2+C_2)}
\end{equation}

where $\mu_x$, $\mu_y $  are the average of x and y respectively,  $\sigma_x^2$, $\sigma_y^2$ are the variance of x and y respectively, $\sigma_{xy}$ is the covariance of x and y, $C_1$, $C_2$ are the constants.

SISR is a low-level image processing task that can be used in a number of contexts. In real-world circumstances, SISR's time constraints are extremely stringent.
We conduct tests to see how long each state-of-the-art method takes to run and then compare the results on Shenzhen Hospital datasets. This is depicted in Figure 13.


%



\ifCLASSOPTIONcompsoc
  \section*{Acknowledgments}

\else
  \section*{Acknowledgment}
\fi

  This work was supported by Natural Science Foundation of China, Grant/Award Number: 61836013; Beijing Natural Science Foundation, Grant/Award Number: 4212030; Beijing Technology, Grant/Award Number:Z191100001119090; Key Research Pr gram of Frontier Sciences, CAS, Grant/Award Number:ZDBS‐LY‐DQC016. 
  
\ifCLASSOPTIONcompsoc
  \section*{Data Availability}

\else
  \section*{Data Availability}
\fi
  
We used the two popular datasets CT-images from Shenzhen Hospital datasets\cite{jaeger2014two} and Public MRI da-taset ({https://sites.google.com/view/calgary-campinas-dataset/home}).

  \section{CONCLUSION}
In this study, we designed a multi-attention GAN framework with wavelet transform and transfer learning for multimodal SISR on medical images. This is the first time a multi-attention GAN has been employed in conjunction with a wavelet transform methodology for medical image super-resolution. We also used transfer leaning, which works with multimodality data that decrease the cost of adding new SR challenging targets like disease classification. We trained our network on the high-resolution DIV2K dataset, then applied transfer learning to train and test medical images on the Shenzhen Hospital datasets of various modalities. Furthermore, we used GANs to train a perceptual loss function that can better super-resolve LR features, resulting in improved perceptual quality of the generated images. Due to the 2D-DWT properties, the reconstructed images are more accurate and have more texture information. The utility of transfer learning is that the pre-trained model is fine-tuned using medical images such as cardiac MR scans and CT scans. In particular, we evaluated our outcomes in terms of visual, quantitative, adversarial, and perceptual loss. The PSNR and SSIM measurements are the preliminary step in evaluating the SR approach, however they are insufficient for real-time applications. In the future, we will study how the proposed approach affects target tasks such as disease detection and classification, as well as minimizing the amount of parameters and flops, which will effectively reduce training and inference time.

\ifCLASSOPTIONcaptionsoff
  \newpage
\fi



%
\bibliographystyle{IEEEtran}
\bibliography{mybibfile}

%

\begin{IEEEbiography}[{\includegraphics[width=1in,height=1.25in,clip,keepaspectratio]{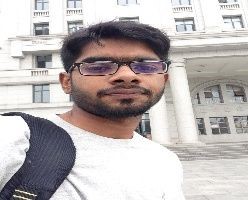}}]{Fayaz Ali Dharejo}
https://orcid.org/0000-0001-7685-391  (ORCID ID).
He received B.E. degree in Electronic Engineering from Quaid-e-Awam University of Engineering Science and Technology (QUEST) Pakistan and M.S. degrees in School of Information and Software Engineering from University of Electronic Science and Technology of China (UESTC) in 2016 and 2018, respectively. Since September 2018, he is Pursuing PhD in the Institute of Computer Network Information Center Chinese Academy of Sciences, University of Chinese Academy of Sciences.. He has published more than 30 articles within few time in reputed ISI impact factor journals and in IEEE conferences, among them the top journals ACM Transactions on Intelligent Systems and Technology, International Journal of Intelligent Systems, IEEEE GRSL and etc. He is a member of professional bodies such as Pakistan Engineering Council (PEC), ACM, IEEE (USA), IEEE Societies, such as IEEE Geoscience and Remote Sensing Society, IEEE Computer Society and IEEE Signal Processing Society.He is a member of the Technical Program Committee and Reviewer of several Scopus, Springer and ISI (Thomson Reuters) indexed conferences.
His research interests are image processing, deep learning, remote sensing and machine learning. 
\end{IEEEbiography}
\begin{IEEEbiography}[{\includegraphics[width=1in,height=1.25in,clip,keepaspectratio]{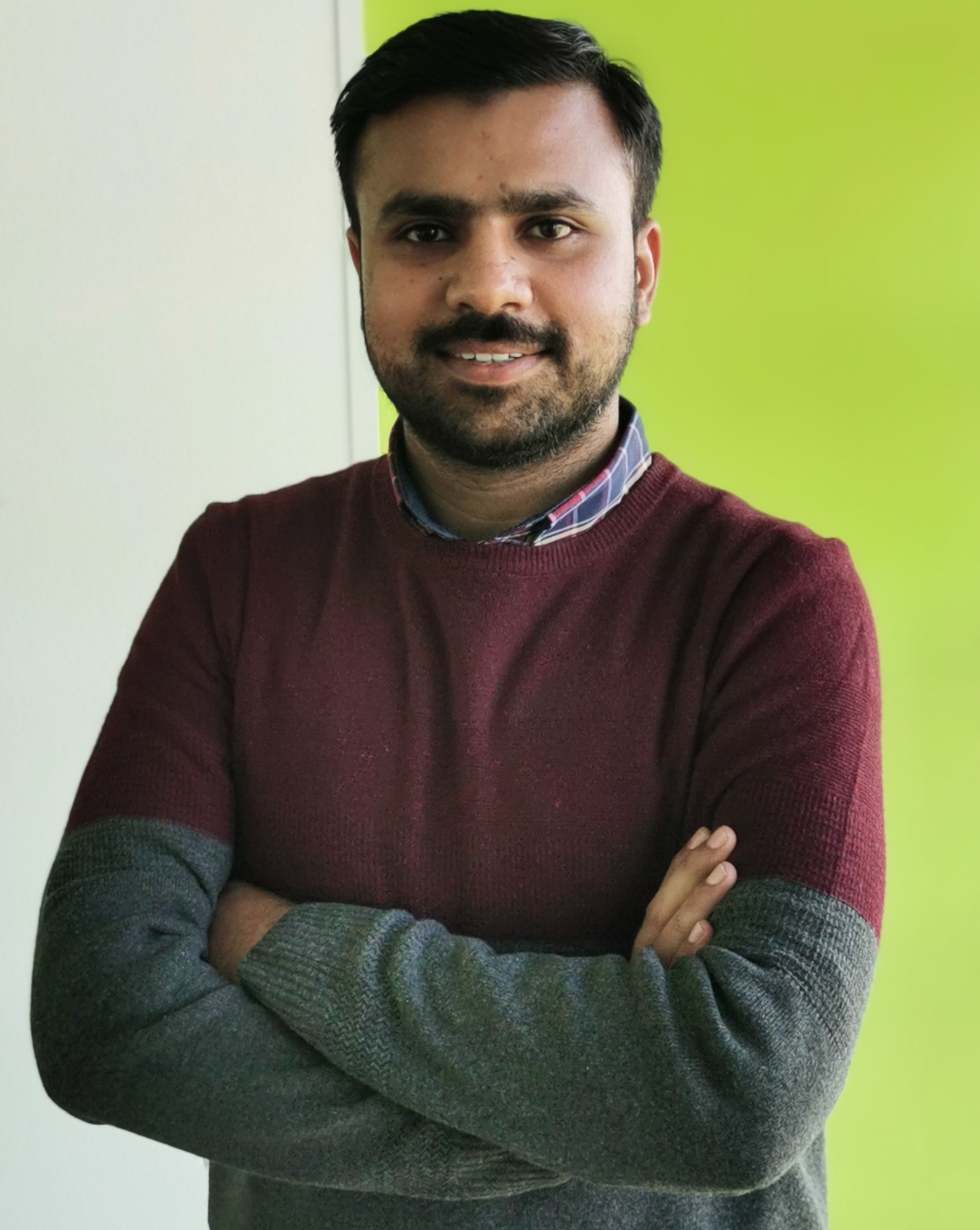}}]{Muhammad Zawish}
has done BE in computer systems from Mehran University of Engineering and Technology, Pakistan. He has worked as Lab engineer in Department of Software Engineering at PAFKIET, Pakistan. Currently, he is pursuing his PhD in computer science at Walton Institute, Waterford Institute of Technology, Ireland. His research interests are deep learning for IoT.
\end{IEEEbiography}
\begin{IEEEbiography}[{\includegraphics[width=1in,height=1.25in,clip,keepaspectratio]{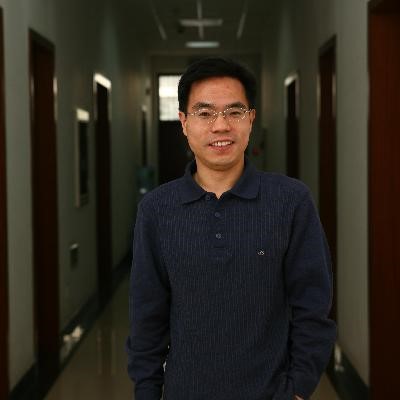}}]{Yuanchun Zhou}
Yuanchun Zhou is a Professor in the Computer Network Infor-mation Center, Chinese Academy of Sciences. He received his PhD from the Institute of Computing Technology, Chinese Academy of Sciences, in 2006. His main research interests in-clude data mining, data intelligence, massive data and data intensive computing. He has published more than 100 research publications.
\end{IEEEbiography}

\begin{IEEEbiography}[{\includegraphics[width=1in,height=1.25in,clip,keepaspectratio]{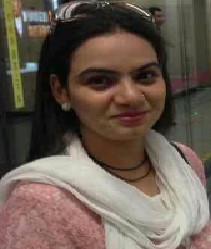}}]{Farah Deeba}
https://orcid.org/0000-0002-1987-2402.
She received B.E. and M.E. degrees in Department of Software Engineering from Mehran University of Engineering and Technology Jamshoro Pakistan in 2010 and 2014, respectively. She has 3 years of experience in teaching at graduate and post-graduate level at Hamdard University Karachi Pakistan. Current-ly, she has completed her Ph.D. degree from the School of Information and Software Engineering, University of Electronic Science and Technology, China in July 2020. Since September 2020 she is also doing Post-doc at Computer network information Center, Chinese Academy of Sciences Beijing, China.  She has served as a reviewer for many ISI (Impact factor) and Scopus indexed journals, Such as IEEE Transactions on Pattern Analysis and Machine Intelligence, Electronics Letters, IET Image Processing, IEEE Access, etc.She is a member of professional bodies such as Pakistan Engineering Council (PEC), ACM, IEEE (USA), IEEE Societies, such as IEEE Geoscience and Remote Sensing Society, IEEE Computer Society and IEEE Signal Processing Society.
Her research interest includes super-resolution, sparse representation and deep learning.
\end{IEEEbiography}
\begin{IEEEbiography}[{\includegraphics[width=1in,height=1.25in,clip,keepaspectratio]{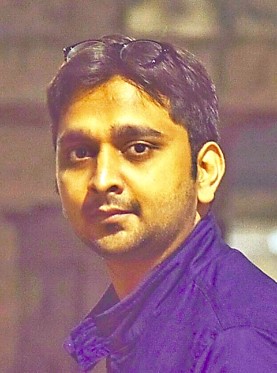}}]{Kapal Dev} ORCID ID: 0000-0003-1262-8594
He is Senior Research Associate at university of Johannesburg South Africa.  Previously, he was a Postdoctoral Research Fellow with the CONNECT Centre, School of Computer Science and Statistics, Trinity College Dublin (TCD). He worked as 5G Junior Consultant and Engineer at Altran Italia S.p.A, Milan, Italy on 5G use cases. He worked as Lecturer at Indus university, Karachi back in 2014. He is also working for OCEANS Network as Head of Projects funded by European Commission. He was awarded the PhD degree by Politecnico di Milano, Italy under the prestigious fellowship of Erasmus Mundus funded by European Commission. His research interests include Blockchain, 6G Networks and Artificial Intelligence. He is very active in leading (as Principle Investigator) Erasmus + International Credit Mobility (ICM), Capacity Building for Higher Education, and H2020 Co-Fund projects. He is also serving as Associate Editor in Springer Wireless Networks, IET Quantum Communication, IET Networks, Topic Editor in MDPI Network, and Review Editor in Frontiers in Communications and Networks. He is also serving as Guest Editor (GE) in several Q1 journals; IEEE TII, IEEE TNSE, IEEE TGCN, Elsevier COMCOM and COMNET. He served as Lead chair in one of IEEE CCNC 2021, IEEE Globecom 2021 and ACM MobiCom 2021 workshops, TPC member of IEEE BCA 2020 in conjunction with AICCSA 2020, ICBC 2021, SSCt 2021, DICG Co-located with Middleware 2020 and FTNCT 2020. He is expert evaluator of MSCA Co-Fund schemes, Elsevier Book proposals and top scientific journals and conferences including IEEE TII, IEEE TITS, IEEE TNSE, IEEE IoT, IEEE JBHI, and elsevier FGCS and COMNET. 
\end{IEEEbiography}
\begin{IEEEbiography}[{\includegraphics[width=1in,height=1.25in,clip,keepaspectratio]{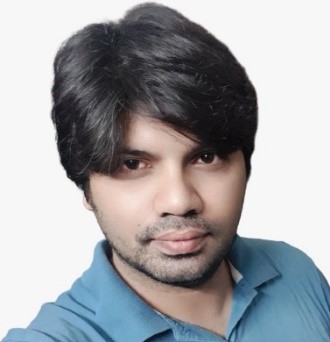}}]{Dr. Sunder Ali Khowaja} ORCID ID: 0000-0002-4586-4131
He received the Ph.D. degree in Industrial and Information Systems Engineering from Hankuk University of Foreign Studies, South Korea. He is currently an Assistant Professor at Department of Telecommunication Engineering, Faculty of Engineering and Technology, University of Sindh, Pakistan. He had the experience of working with multi-national companies as Network and RF Engineer from 2008-2011. He is having teaching, research and administrative experience of more than 10 years. He has published over 30 research articles in national, international journals, and conference proceedings. He is also a regular reviewer of notable journals including IEEE Transactions on Industrial Informatics, IEEE Internet of Things Journal, IEEE Access, Electronic Letters, IET Image Processing, IET Signal Processing, Computer Communications, International Journal of Imaging Systems and Technology, Multimedia Systems, IET Networks, SN Applied Sciences, Artificial Intelligence Review, Computational and Mathematical in Medicine, Computers in Human Behavior, IET Wireless Sensor Systems, Journal of Information Processing Systems, and Mathematical Problems in Engineering. He has also served as a TPC member for workshops in A* conferences such as Globecom and Mobicom. His research interests include Data Analytics, Process Mining, and Deep Learning for Computer Vision applications and Emerging Communication Technologies. 
\end{IEEEbiography}
\begin{IEEEbiography}[{\includegraphics[width=1in,height=1.25in,clip,keepaspectratio]{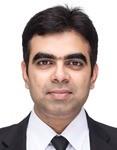}}]{Dr. Nawab Muhammad Faseeh Qureshi} ORCID ID: 0000-0002-5035-2640
 He is an Assistant Professor at Sungkyunkwan University, Seoul, South Korea. He received Ph. D. in Computer Engineering from Sungkyunkwan University, South Korea, through a SAMSUNG scholarship. He was awarded the 1st Superior Research Award from the College of Information and Communication Engineering based on his research contributions and performance during his studies. He has served as Guest Editor in more than 17 Journals. Also, he has served as General Chair Workshop in many conferences that includes IEEE Wireless Communications and Networking Conference (WCNC2020) and IEEE Globecom 2020.  In addition to that, he has served as Proceedings Chair in Global Conference on Wireless and Optical Technologies consecutively for the third time. He is a reviewer of various prestigious IEEE and ACM Journals. He serves in many leading IEEE and ACM conferences as a reviewer every year. He has evaluated several theses as external Ph.D. thesis evaluators. Also, he has facilitated several institutes and conferences with keynotes and Webinars on Big data analysis and Modern Technology convergence. He is an active Senior Member of IEEE, ACM, KSII (Korean Society for Internet Information), and IEICE (Institute of Electronics, Information and Communication Engineers). His research interests include big data analytics, machine learning, deep learning, context-aware data processing of the Internet of Things, and cloud computing.

\end{IEEEbiography}





\end{document}